\documentclass[aps,prl,twocolumn]{revtex4}
\usepackage{amssymb}
\usepackage{amsmath}
\usepackage{graphicx}
\usepackage{nicefrac}

\begin{document}

\title{Near doping-independent pocket area from an antinodal Fermi surface instability in underdoped high temperature superconductors}
\author{N.~Harrison%$^1$, S.~E.~Sebastian$^2$
}
\affiliation{%$^1$
Los Alamos National Laboratory, MS E536, Los Alamos, New Mexico 87545\\
%$^2$Cavendish Laboratory, Cambridge University, JJ Thomson Avenue, Cambridge CB3~OHE, U.K
}
\date{\today}

\begin{abstract}
Fermi surface models applied to the underdoped cuprates predict the small pocket area to be strongly dependent on doping whereas quantum oscillations in YBa$_2$Cu$_3$O$_{6+x}$ find precisely the opposite to be true $-$ seemingly at odds with the Luttinger volume. We show that such behavior can be explained by an incommensurate antinodal Fermi surface nesting-type instability $-$ further explaining the doping-dependent superstructures seen in cuprates using scanning tunneling microscopy. We develop a Fermi surface reconstruction scheme involving orthogonal density waves in two-dimensions and show that their incommensurate behavior requires momentum-dependent coupling. A co-operative modulation of the charge and bond-strength is therefore suggested.
\end{abstract}

\maketitle

\begin{figure}[htbp!]
\centering
%\vspace{-2mm}
\includegraphics[width=0.45\textwidth]{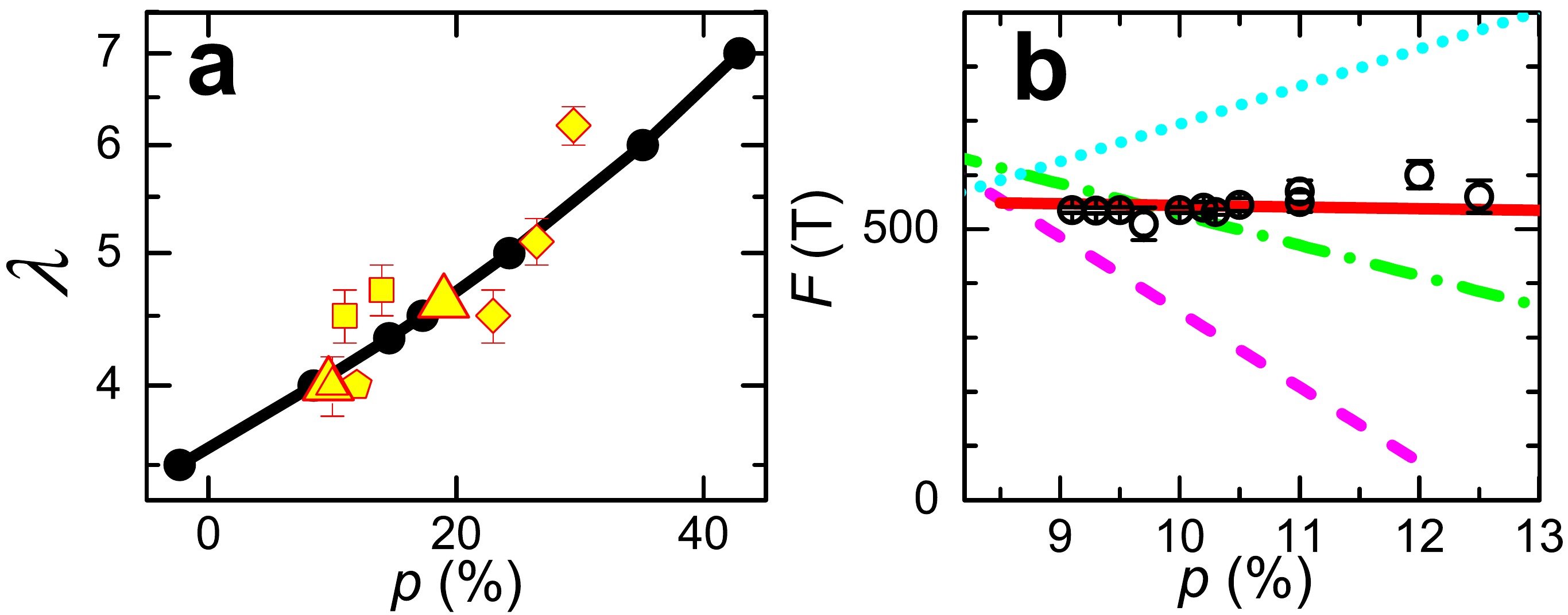}
%\vspace{-2mm}
\caption{{\bf a} Charge modulation periods seen using x-rays and NQR in YBa$_2$Cu$_3$O$_{6+x}$~\cite{liu1,julien1} (large triangles), neutrons in La$_{0.4}$Nd$_{0.4}$Sr$_{0.12}$CuO$_4$~\cite{tranquada1} (pentagon) and STM in Bi$_{2-y}$Pb$_y$Sr$_{2-z}$La$_z$CuO$_{6+x}$ (diamonds), Bi$_2$Sr$_2$CaCu$_2$O$_{8+\delta}$ (squares) and Ca$_{2-x}$Na$_x$CuO$_2$Cl$_2$ (small triangle) taken from Ref.~\cite{wise1}. In comparing different materials, we neglect possible differences in $\varepsilon({\bf k})$~\cite{band}. The line and circles show the $p$ for each $\lambda$ extracted from the model density-of-states minimum (e.g. Fig.~\ref{dos}b). {\bf b} Measured leading YBa$_2$Cu$_3$O$_{6+x}$ quantum oscillation frequency~\cite{sebastian1,vignolle1, sebastian3,singleton1,phase} (circles) compared to its strong $p$-dependence expected in the 4 hole pocket~\cite{lee1} (dotted line), Millis and Norman stripe~\cite{millis1} (dot-dash line) and fixed $\lambda=4$ bidirectional charge~\cite{harrison1} (dashed line) models, where $F=(\hbar/2\pi e)A_{\rm e}$. The present model (solid line) uniquely yields a weakly $p$-dependent $F$~\cite{phase}.
}
\label{doping}
%\vspace{-3mm}
\end{figure}

Identification of the forms of order competing with superconductivity and antiferromagnetism in the high-$T_{\rm c}$ cuprates remains a considerable experimental challenge~\cite{lee1,broun1}. Among possibilities, charge ordering is reported in several experiments within the underdoped regime $-$ namely x-ray diffraction~\cite{liu1}, neutron scattering~\cite{tranquada1}, scanning tunneling microscopy (STM)~\cite{hoffman1} and nuclear quadrupole resonance (NQR)~\cite{julien1} (see Fig.~\ref{doping}{\bf a}). Yet its extent and relevance are far from understood. It is yet to be established whether such ordering participates in forming the pseudogap~\cite{hoffman1}, whether it is inherently unidirectional as opposed to bidirectional in nature~\cite{kivelson1}, or whether it is caused by a Fermi surface instability~\cite{wise1} as opposed to being a biproduct of spin order~\cite{tranquada1}.  

In the light of recent quantum oscillation~\cite{doiron1,yelland1}, electrical transport~\cite{leboeuf1} and angle-resolved photoemission spectroscopy (ARPES)~\cite{hashimoto1} studies, several Fermi surface reconstruction models have been invoked in the underdoped cuprates postulating charge (and/or other forms of) ordering~\cite{millis1,chakravarty1,yelland1,harrison1}. A serious problem with {\it all} proposed models, however, is that they predict the pocket size to be strongly dependent on the hole doping (e.g. dotted and dashed lines in Fig.~\ref{doping}{\bf b}), whereas experiments on underdoped YBa$_2$Cu$_3$O$_{6+x}$~\cite{sebastian1,vignolle1, sebastian3,singleton1} find the pocket area to change remarkably little over a range of hole dopings spanning $\approx$~3~\%~\cite{phase} (circles in Fig.~\ref{doping}{\bf b}). 

\begin{figure}[htbp!]
\centering
%\vspace{-8mm}
\includegraphics[width=0.45\textwidth]{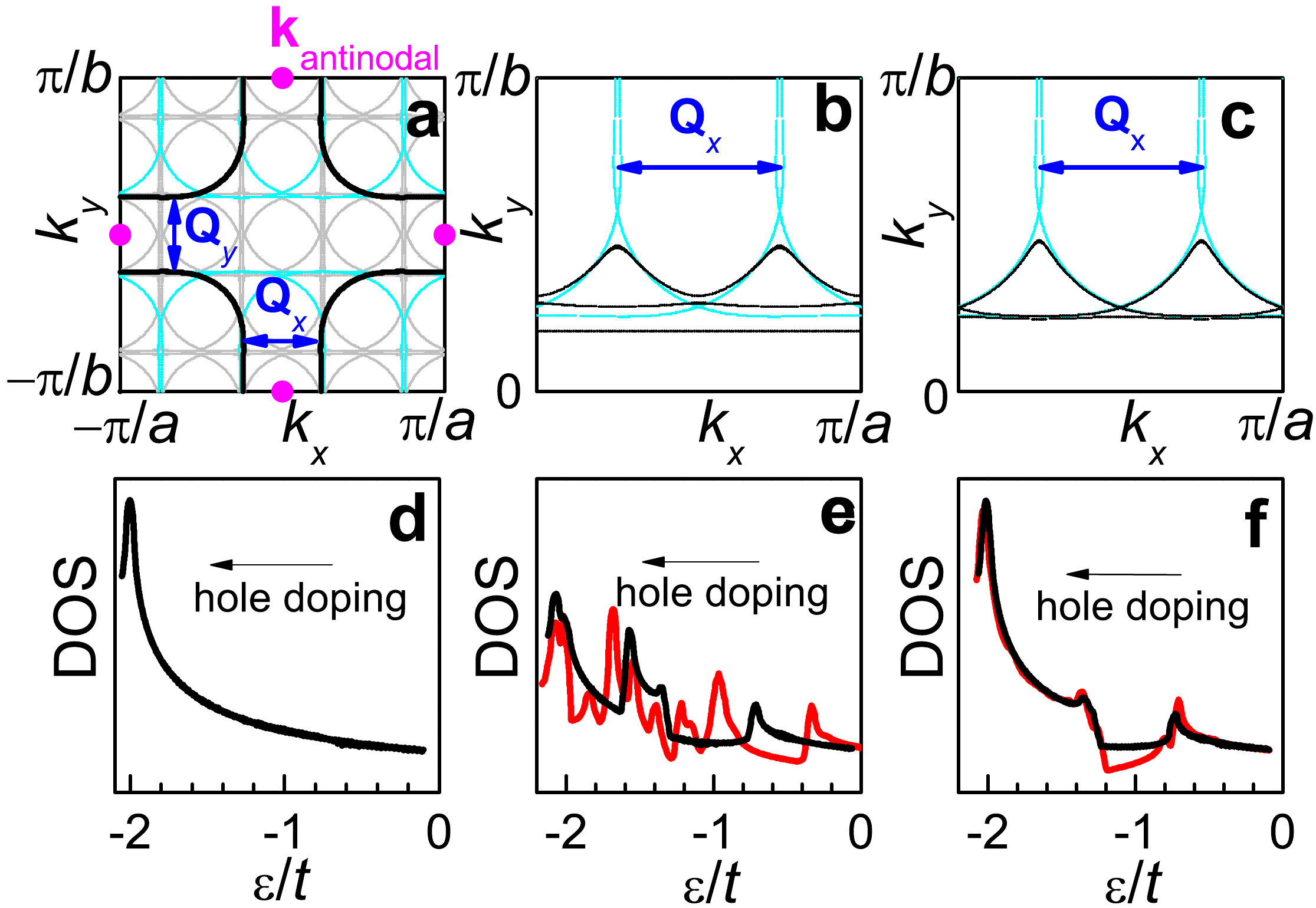}
%\vspace{-7mm}
\caption{{\bf a} The unreconstructed Fermi surface at $p=$~8.5~\% (black)~\cite{band} together with itself translated by multiples of ${\bf Q}_x$ (cyan) and multiples of ${\bf Q}_x$ and ${\bf Q}_y$ (grey) for $\lambda=4$. {\bf b} Reconstructed Fermi surface (black) resulting from a unidirectional charge modulation (i.e. $n^\prime=$~1) in which $V_{x,0}=0.3t$ and $r=0$ in Equation~(\ref{nestingselective}), shown for a quadrant of the extended Brillouin zone. {\bf c} Same as ({\bf b}) but with a momentum-dependent $V_x$ in which we choose $r=1$~\cite{currents}. {\bf d} The calculated electronic density-of-states (DOS) for the unreconstructed band~\cite{band} exhibiting a van-Hove singularity near $\varepsilon\approx-2t$. {\bf e} The calculated DOS (black line) for $r=0$ in ({\bf b}). {\bf f} The calculated DOS (black line) for $r=1$ in ({\bf c}). Red lines in ({\bf e}) and ({\bf f}) are the corresponding DOS calculated for concurrent charge modulations along $a$ and $b$ (i.e. such that $n^\prime=n$) in which we assume $V_{x,0}=V_{y,0}$ (by no means a required constraint).}
\label{simple}
%\vspace{-3mm}
\end{figure}

In this paper, we show that the near doping-independence of the orbit area in underdoped YBa$_2$Cu$_3$O$_{6+x}$~\cite{vignolle1,sebastian1,phase} and the increasing charge modulation period seen with hole doping in STM experiments on Bi$_{2-y}$Pb$_y$Sr$_{2-z}$La$_z$CuO$_{6+x}$~\cite{wise1} can both be consistently explained by Fermi surface reconstruction resulting from an antinodal Fermi surface nesting-type instability (i.e. at $[\pm\frac{\pi}{a},0]$ and $[0,\pm\frac{\pi}{b}]$ in Fig.~\ref{simple}{\bf a}). We present a density wave model in which we mimic incommensurate behavior by considering modulation periods $\lambda$ corresponding to different rational multiples of the in plane lattice vectors (e.g. $\lambda=$~$\nicefrac{7}{2}$, 4, $\nicefrac{13}{3}$, $\nicefrac{9}{2}$, 5, 6 and 7). On treating scenarios in which the coupling between translated bands is uniform (as in a charge-density wave~\cite{millis1,harrison1}) or acquires a momentum-dependence (as occurs on incorporating a bond-density wave component~\cite{nayak1}), we find that only the latter leads to a single well-defined gap in the electronic density-of-states at weak couplings $V_{x,y}\ll t_{10}$ (where $t_{10}$ is the nearest neighbor hopping~\cite{band}). We show the latter also to be a necessary prerequisite for incommensurate behavior, in which the electronic structure evolves continuously as a function of $\lambda$. 

We model Fermi surface reconstruction caused by modulations of general period $\lambda=\nicefrac{n}{m}$ (in which $n$ and $m$ are integers) along the $a$ and/or $b$ lattice directions by diagonalizing a Hamiltonian consisting of nested matrices
\begin{widetext}
\begin{equation}\label{matrix1}
{\bf H}_{xy}=\left( \begin{array}{cccccc}
{\bf H}_x(0) & V_y{\bf I}_n & 0 &\dots & 0& V_y{\bf I}_n\\
V_y{\bf I}_n & {\bf H}_x(1) & V_y{\bf I}_n &\dots&0& 0\\
0 & V_y{\bf I}_n &{\bf H}_x(2) &\dots&0&0\\
\vdots & \vdots & \vdots &\ddots&\vdots&\vdots\\
0 & 0 & 0&\dots&{\bf H}_x(n^\prime-2)&V_y{\bf I}_n\\
V_y{\bf I}_n & 0 & 0 &\dots&V_y{\bf I}_n& {\bf H}_x(n^\prime-1)\\ \end{array} \right).
\end{equation}
Here, ${\bf I}_n$ is an identity matrix of rank $n$, $n^\prime=n$ for bidirectional order (or $n^\prime=1$ for unidirectional order),
\[
{\bf H}_x(i)=\left( \begin{array}{cccccc}
\varepsilon_{i{\bf Q}_y} & V_x & 0 &\dots & 0& V_x\\
V_x & \varepsilon_{{\bf Q}_x+i{\bf Q}_y} & V_x &\dots&0& 0\\
0 & V_x & \varepsilon_{2{\bf Q}_x+i{\bf Q}_y} &\dots&0&0\\
\vdots & \vdots & \vdots &\ddots&\vdots&\vdots\\
0 & 0 & 0&\dots&\varepsilon_{(n-2){\bf Q}_x+i{\bf Q}_y}&V_x\\
V_x & 0 & 0 &\dots&V_x& \varepsilon_{(n-1){\bf Q}_x+i{\bf Q}_y}\\ \end{array} \right)\]
\end{widetext}
and $\varepsilon_{j{\bf Q}_x+i{\bf Q}_y}$ represents the electronic dispersion $\varepsilon({\bf k})$~\cite{band} subject to translation by multiples of ${\bf Q}_x=[\frac{2\pi}{\lambda a},0]$ and ${\bf Q}_y=[0,\frac{2\pi}{\lambda b}]$.  

In the case of a conventional density wave, the normal assumption is for the potentials to uniformly couple all band crossings subject to a relative translation by ${\bf Q}_x$ or ${\bf Q}_y$ such that $V_x=V_{x,0}$ and $V_y=V_{y,0}$ are constants in Equation (\ref{matrix1}). In the case of incommensurate ordering in a two-dimensional lattice, however, the coupling $V$ has been found to vary depending on the band crossing in question~\cite{brouet1,rossnagel1}. Such behavior is most apparent in $R$Te$_3$~\cite{brouet1} (owing to its exceptionally large gap), where ARPES finds a momentum-dependent $V({\bf k})$ that selectively couples portions of the Fermi surface subject to nesting.

While the real-space implications of a momentum-dependent coupling in the chalcogenides has yet to be investigated, in the cuprates it is connected with the possibility of bond-strength or bond-current density wave ordering~\cite{nayak1}. We find a simple form of the coupling~\cite{diagonal},
\begin{eqnarray}\label{nestingselective}
V_x({\bf k})=V_{x,0}\tfrac{1}{1-r}(1-r\cos bk_y)~\nonumber\\V_y({\bf k})=V_{y,0}\tfrac{1}{1-r}(1-r\cos bk_x),
\end{eqnarray}
in which $r$ adds a bond-strength modulation to an otherwise conventional charge-density wave, to prove particularly effective at reducing the electronic density-of states (and consequent free energy) when $r\approx$~1~\cite{currents}. It does so by suppressing $V({\bf k})$ in the regions of the Brillouin zone where unnested bands cross~\cite{currents}, which we demonstrate in Fig.~\ref{simple} by considering the simple case of a unidirectional modulation ${\bf Q}_x=[\frac{2\pi}{\lambda a},0]$ [in which $\lambda=$~4, $n=4$ and $n^\prime=$~1 in Equation (\ref{matrix1})].

From Figs.~\ref{simple}{\bf b} and {\bf c} it is evident that while both uniform ($r=0$) and strongly momentum-dependent ($r=1$) forms of $V_x$ open a gap at $|k_y|>\frac{\pi}{2b}$, where the Fermi surfaces are nested by ${\bf Q}_x$, the latter does so without splitting the open Fermi surfaces at $k_y\approx\pm\frac{\pi}{4b}$. The splitting in Fig.~\ref{simple}{\bf b} occurs concomitantly with an additional gap in the electronic density-of-states at $\varepsilon\approx$~-1.8~$t$ in Fig.~\ref{simple}{\bf e} and a slightly weaker ordering gap at the Fermi energy ($\varepsilon_{\rm F}\approx-t$) than in Fig.~\ref{simple}{\bf f}. A large $V_x$ at $|k_y|\approx\frac{\pi}{4b}$ is therefore energetically unfavorable~\cite{currents}. The momentum-dependent $V_x$ (i.e. $r\approx1$) avoids unfavorable splittings and gaps, leaving the remaining open Fermi surfaces at $k_y\approx\pm\frac{\pi}{4b}$ amenable to a secondary Fermi surface instability of wavevector ${\bf Q}_y=[0,\frac{2\pi}{\lambda b}]$, which can further lower  the density-of-states (and consequently the electronic energy) by forming a concurrent modulation along $b$ (red line in Fig.~\ref{simple}{\bf f}) [where $n^\prime=n=4$ in Equation~(\ref{matrix1}) in the case of bidirectional ordering]. By contrast, the splittings caused by a uniform potential (i.e. $r=0$) mutually disrupt nesting for both ${\bf Q}_x$ and ${\bf Q}_y$ in the case of bidirectional ordering, leading to an energetically unfavorable higher density-of-states consisting of multiple peaks and valleys in the vicinity of the Fermi energy (red line in Fig.~\ref{simple}{\bf e}).

\begin{figure}[htbp!]
\centering
%\vspace{-8mm}
\includegraphics[width=0.45\textwidth]{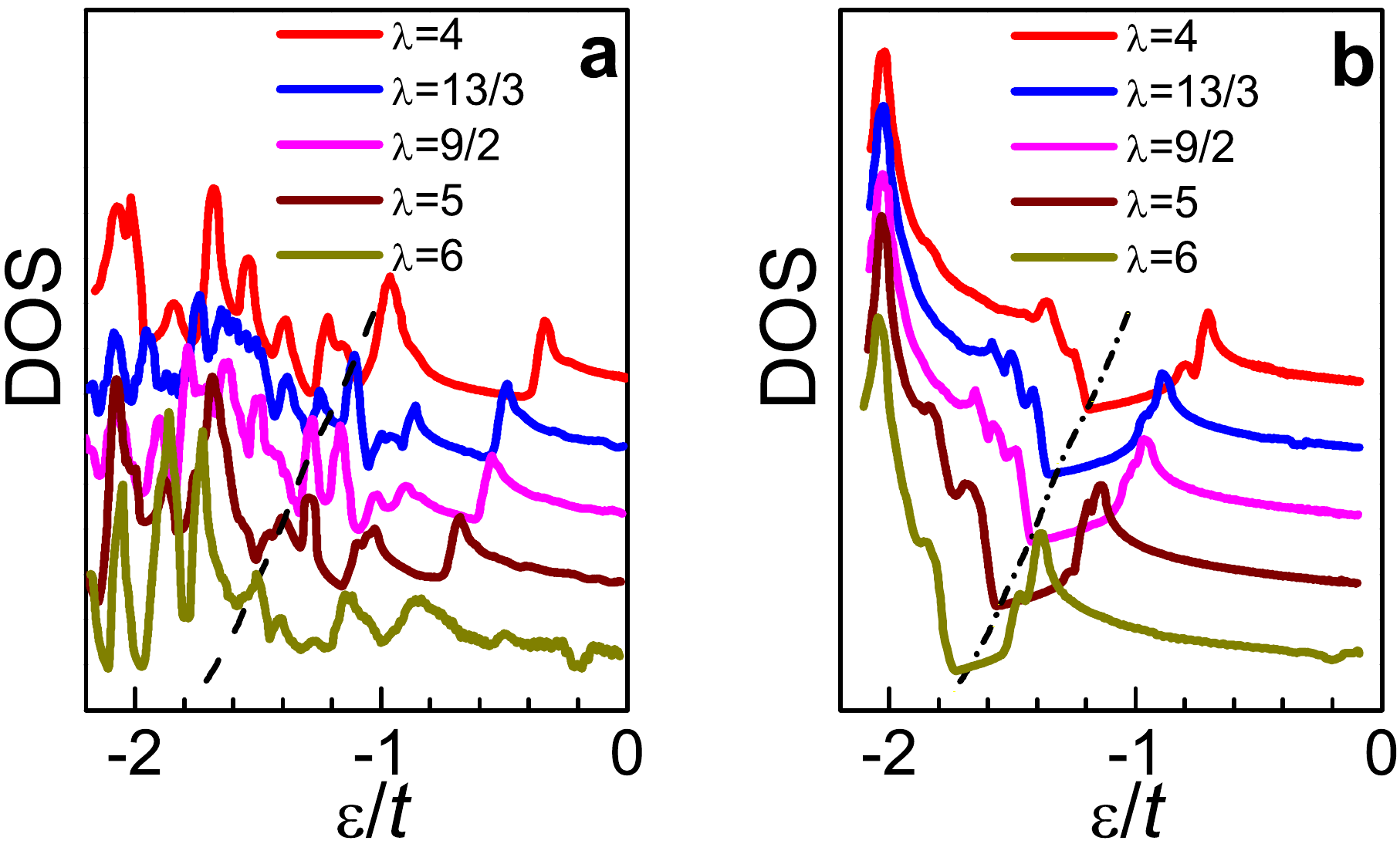}
%\vspace{-7mm}
\caption{{\bf a} Electronic density-of-states (DOS) calculated for bidirectional charge modulations with periodicities corresponding to different multiples $\lambda$ of the $a$ and $b$ lattice vectors as indicated, assuming uniform couplings $V_{x,0}=V_{y,0}=0.3t$ in which $r=0$. {\bf b} Same as ({\bf b}) but assuming momentum-dependent couplings in which $r=1$ in Equation~(\ref{nestingselective}). Curves have been offset for clarity. The dashed line in ({\bf b}) indicates the minimum in the DOS near to which the Fermi energy is likely to be located.
}
\label{dos}
%\vspace{-3mm}s
\end{figure}

On extending the bidirectional ordering density-of-states calculation to different periods in Fig.~\ref{dos}, we continue to find a well defined single gap with a broad deep minimum {\it only} in the case of momentum-dependent coupling (see Fig.~\ref{dos}{\bf b}), pointing to its continuous evolution with $\lambda$. In the case of a uniform coupling (see Fig.~\ref{dos}{\bf a}), by contrast, the multiple peaks and valleys vary discontinuously with $\lambda$. 

\begin{figure}[htbp!]
\centering
%\vspace{-8mm}
\includegraphics[width=0.45\textwidth]{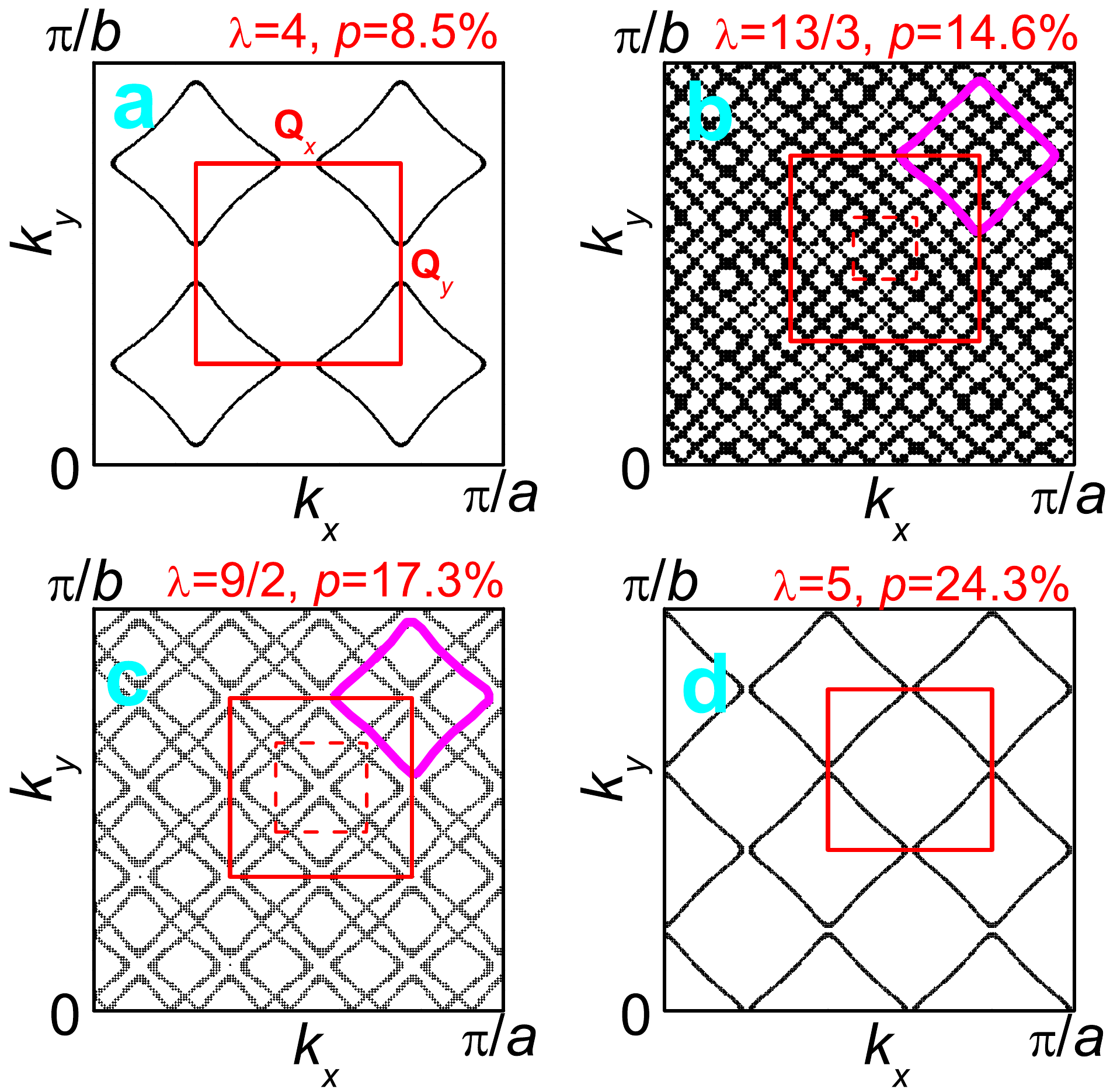}
%\vspace{-7mm}
\caption{{\bf a}, {\bf b}, {\bf c} and {\bf d} Reconstructed Fermi surface for selected $\lambda$'s in Fig.~\ref{dos}{\bf b} when the Fermi energy is situated at the minimum in the density-of-states, with the corresponding hole doping given.  Solid red lines indicate the $k$-space area of the $\lambda a\times\lambda b$ superstructure, while dashed red lines indicate the $n^2$-fold reduced Brillouin zone [which coincides with the superstructure in ({\bf a}) and ({\bf d})]. In ({\bf b}) and ({\bf c}), a magenta line is used to trace the path of the electron-like orbit that occurs in strong magnetic fields. $t_{10}\sim$~100~meV~\cite{band} produces an effective mass consistent with experiments.}
\label{FS}
%\vspace{-3mm}
\end{figure}

Thus, by generating a deep wide gap in the density-of-states whose form and location in energy shifts continuously with $\lambda$, momentum-dependent coupling provides an incentive for incommensurate behavior in which $\lambda$ adjusts itself in a continuous fashion so as to lower the electronic energy. Because the electronic energy in an itinerant picture is minimized by having the Fermi energy situated within a broad deep gap in the density-of-states, the $\lambda$-dependent gap provides an explanation for the evolution of the periodic structures seen in STM experiments as a function of doping~\cite{wise1}. The location of the minimum (identified by the dot-dashed line in Fig.~\ref{dos}{\bf b}) enables us to estimate the hole doping $p$ at which each period is most likely to be stable (plotted in Fig.~\ref{doping}{\bf a}). Using these dopings and assuming Luttinger's theorem~\cite{luttinger1}, we calculate the corresponding Fermi surfaces in Fig.~\ref{FS}, whose forms consist of a single electron orbit (located close to the nodes) consistent with experimental observations~\cite{harrison1,sebastian2}. Momentum-dependent coupling enables such a pocket to exist for weaker couplings than in ref.~\cite{harrison1} and to persist essentially unchanged as a function of doping. Most importantly, the near $p$-independent area (solid line in Fig.~\ref{doping}{\bf b}) reproduces experimental observations (circles). 

The sub-gaps occurring at the intersections of the electron orbits in Figs.~\ref{FS}{\bf b} and {\bf c} are small enough [$\Delta^2_{\rm sub}/BF(\hbar e/m^\ast)^2\ll1$ provided $V_{x,y}\ll t_{10}$] to be completely broken down~\cite{lomer1} in magnetic fields of the strength required to see magnetic quantum oscillations~\cite{doiron1,yelland1,leboeuf1,sebastian1,vignolle1,sebastian2} $-$ giving rise to a single orbit (thick magenta line) in strong magnetic fields. The sub-gaps nevertheless imply the absence of a simple ($\lambda$-independent) relationship between the quantum oscillation frequency $F_{\rm e}=(\hbar/2\pi e)A_{\rm e}$ and the frequency $F_{\rm L}=\frac{p}{2}F_{\rm BZ}$ corresponding to the Luttinger hole doping (where $F_{\rm BZ}=h/eab$ is the unreconstructed Brillouin zone frequency). Only when the density-wave is `accidentally' commensurate such that sub-gaps do not occur (e.g. $\lambda=$~4 or 5) can adherence to Luttinger's theorem~\cite{luttinger1} be easily verified in quantum oscillation experiments. In Fig.~\ref{FS}{\bf a}, for example, $F_{\rm L}=F_\lambda-F_{\rm e}$ (where $F_\lambda=F_{\rm BZ}/\lambda^2$ is the $\lambda a\times\lambda b$ superstructure frequencsy), while in Fig.~\ref{FS}{\bf d} it is given by $F_{\rm L}=\frac{7}{2}F_\lambda-F_{\rm e}$.

Finally, we turn to aspects of momentum-selective density waves that may potentially be reconciled with the unidirectional behavior of charge ordering noted in the cuprates~\cite{kivelson1}. While closed Fermi surface pockets require charge modulations to occur concurrently along $a$ and $b$ lattice directions (in the absence of other orders~\cite{harrison1}), the superposition of their ordering gaps in Fig.~\ref{simple}{\bf f} (red line) implies the absence of a significant energy penalty (or interaction) associated with their coexistence $-$ in contrast to the uniform case in Fig.~\ref{simple}{\bf e} where such a superposition does not occur. Given the implied independence of the modulations along $a$ and $b$, underlying anisotropies in the electronic structure (such as that caused by the presence of chains in YBa$_2$Cu$_3$O$_{6+x}$~\cite{andersen1}) will likely produce anisotropies in $V_{x,y,0}$, $\lambda$, $r$ and the onset temperature. In the present simulations, we find a Fermi surface consisting solely of an electron pocket to remain robust against an anisotropy $V_{x,0}/V_{y,0}$ as large as 4.

In conclusion, we present a model that explains the lack of a detectable doping-dependence of the quantum oscillation frequency in underdoped YBa$_2$Cu$_3$O$_{6+x}$ (i.e. Fig.~\ref{doping}{\bf b}~\cite{phase}). By considering rational values of $\lambda$, we develop what is in essence an incommensurate model for co-operative charge- and bond-density wave ordering in the cuprates~\cite{nayak1,sachdev1} $-$ here driven by a Fermi surface instability at the antinodes. By incorporating a (possibly dominant~\cite{currents}) bond-density wave component~\cite{nayak1}, the size of the periodic potential required to produce a single pocket with a small residual density-of-states is greatly reduced (i.e. $V_{x,y,0}\gtrsim0.05t_{10}$)~\cite{currents} relative to other models~\cite{millis1,chakravarty1,harrison1}. A key strength of the present model is its ability to reconcile doping-dependent quantum oscillations~\cite{sebastian1,vignolle1,singleton1,phase,sebastian2,sebastian3} with the doping-dependent $\lambda$ seen in STM and other experiments~\cite{liu1,julien1,wise1} (i.e. Fig.~\ref{doping}{\bf b}), the negative Hall and Seebeck coefficients over a broad range of dopings seen in transport~\cite{leboeuf1} and particle-hole symmetry breaking reported at the antinodes in ARPES~\cite{hashimoto1, broun1} $-$ all while maintaining compliance with Luttinger's theorem~\cite{luttinger1}.

The author acknowledges the DOE BES project ``Science at 100 Tesla'' and useful discussions with Ross McDonald and Arkady Shehter.


\begin{thebibliography}{99}
\bibitem{phase} The scatter is related to small number of oscillations at some dopings. Upon close examination, the resistive oscillations of YBa$_2$Cu$_3$O$_{6+x}$ in Refs.~\cite{sebastian1,sebastian2,sebastian3,vignolle1} are seen to remain essentially in-phase for the range of dopings spanning $\approx$~3~\%, implying that the same Landau level crosses the Fermi energy at the same field to within $\Delta B\sim$~0.5~T in all samples regardless of doping. From this we infer $F$ to change by less than $(\Delta B/B)F\sim$~5~T (corresponding to $<$~1~\%~change in orbit area $A_{\rm e}$).
\bibitem{band} We use
$\varepsilon=\varepsilon_0+2t_{10}[\cos ak_x+\cos bk_y]+2t_{11}[\cos(ak_x+bk_y)+\cos(ak_x-bk_y)]+2t_{20}[\cos 2ak_x+\cos 2bk_y]$~ \cite{andersen1}, with $t_{11}/t_{10}=-$~0.5 and $t_{20}/t_{10}=$~0.1 chosen to reproduce the YBa$_2$Cu$_3$O$_{6+x}$ Fermi surface from recent ARPES~\cite{elfimov1}.
\end{thebibliography}
\end{document}